\documentclass[runningheads,table]{llncs}
\usepackage{graphicx}
\usepackage[numbers,sort&compress,sectionbib]{natbib}
\usepackage{amsmath,graphicx}
\usepackage{amsfonts}
\usepackage{acro}
\usepackage{tabularx}
\graphicspath{{Figure/}}
\usepackage{color}
\usepackage{multirow}
\usepackage{siunitx}
\usepackage{booktabs} 
\usepackage[font={small}]{caption, subfig}
\usepackage{tikz}

\usepackage{arydshln}
\usepackage{multirow}
\usepackage[utf8]{inputenc}
\usepackage{array}

\usepackage{epstopdf}
\usepackage{ulem}
\usepackage[hyperfootnotes=false]{hyperref}

\newcolumntype{Y}{>{\centering\arraybackslash}X}

\newcommand{\Fig}{Fig.}

\AtBeginDocument{}

\newcommand\blfootnote[1]{%
  \begingroup
  \renewcommand\thefootnote{}\footnote{#1}%
  \addtocounter{footnote}{-1}%
  \endgroup
}

\DeclareAcronym{LNS}{
short=LNS,
long= lymph node station
}

\DeclareAcronym{LN}{
short=LN,
long= lymph node
}

\DeclareAcronym{CT}{
short=CT,
long= computed tomography
}

\DeclareAcronym{IASLC}{
short=IASLC,
long= International Association for the Study of Lung Cancer
}

\DeclareAcronym{NAS}{
short=NAS,
long= neural architecture search 
}

\DeclareAcronym{CE}{
short=CE,
long= contrast enhanced
}

\DeclareAcronym{CNN}{
short=CNN,
long= convolutional neural network
}

\DeclareAcronym{DS}{
short=DSC,
long= Dice score
}

\DeclareAcronym{ASD}{
short=ASD,
long= average surface distance
}

\DeclareAcronym{HD}{
short=HD,
long= Hausdorff distance
}

\begin{document}

\title{DeepStationing: Thoracic Lymph Node Station Parsing in CT Scans using Anatomical Context Encoding and Key Organ Auto-Search}
\author{Dazhou Guo\textsuperscript{*}\textsuperscript{1}  \and 
Xianghua Ye\textsuperscript{*}\textsuperscript{2} \and 
Jia Ge\textsuperscript{2} \and 
Xing Di\textsuperscript{3} \and 
Le Lu\textsuperscript{1} \and 
Lingyun Huang\textsuperscript{4} \and 
Guotong Xie\textsuperscript{4} \and 
Jing Xiao\textsuperscript{4} \and 
Zhongjie Lu\textsuperscript{2} \and 
Ling Peng\textsuperscript{5} \and 
Senxiang Yan\textsuperscript{2} \and 
Dakai Jin\textsuperscript{1}}
\institute{\textsuperscript{1}PAII Inc., Bethesda, MD, USA\\ \textsuperscript{2}The First Affiliated Hospital Zhejiang University, Hangzhou, China\\ \textsuperscript{3}Johns Hopkins University, Baltimore, USA\\ \textsuperscript{4} Ping An Insurance Company of China, Shenzhen, China\\ \textsuperscript{5} Zhejiang Provincial People's Hospital, Hangzhou, China \\
\footnotesize \{guo2004131, dakai.jin\}@gmail.com, \{hye1982, yansenxiang\}@zju.edu.cn }

\authorrunning{D. Guo et al.}
\titlerunning{DeepStationing: Thoracic Lymph Node Station Parsing in CT}
%
%
%
%
\maketitle              

\begin{abstract}
\Ac{LNS} delineation from \ac{CT} scans is an indispensable step in radiation oncology workflow. High inter-user variabilities across oncologists and prohibitive laboring costs motivated the automated approach. Previous works exploit anatomical priors to infer \ac{LNS} based on predefined ad-hoc margins. However, without the voxel-level supervision, the performance is severely limited. \ac{LNS} is highly context-dependent---LNS boundaries are constrained by anatomical organs---we formulate it as a deep spatial and contextual parsing problem via encoded anatomical organs. This permits the deep network to better learn from both CT appearance and organ context. We develop a stratified referencing organ segmentation protocol that divides the organs into anchor and non-anchor categories and uses the former's predictions to guide the later segmentation. We further develop an auto-search module to identify the key organs that opt for the optimal \ac{LNS} parsing performance. Extensive four-fold cross-validation experiments on a dataset of $98$ esophageal cancer patients (with the most comprehensive set of  \textit{12 \acp{LNS} + 22 organs} in thoracic region to date) are conducted. Our LNS parsing model produces significant performance improvements, with an average Dice score of $81.1\%\pm6.1\%$, which is $5.0\%$ and $19.2\%$ higher over the pure CT-based deep model and the previous representative approach, respectively. \blfootnote{* equal contribution.}


\end{abstract}

\acresetall
\section{Introduction}

Cancers in thoracic region are the most common cancers worldwide~\cite{sung2021global} and significant proportions of patients are diagnosed at late stages involved with lymph node (LN) metastasis. The treatment protocol is a sophisticated combination of surgical resection and chemotherapy and/or radiotherapy~\cite{hirsch2017lung}. Assessment of involved LNs~\cite{zhu2020lymph,chao2020lymph} and accurate labeling their corresponding stations are essential for the treatment selection and planning. For example, in radiation therapy, the delineation accuracy of gross tumor volume (GTV) and clinical target volume (CTV) are the two most critical factors impacting the patient outcome. For CTV delineation, areas containing metastasis \acp{LN} should be included to sufficiently cover the sub-clinical disease regions~\cite{chapet2005ct}. One strategy to outline the sub-clinical disease region is to include the \ac{LNS} that containing the metastasized \acp{LN}~\cite{pignon1992meta,yuan2019lymph}. Thoracic LNS is determined according to the text definitions of \ac{IASLC}~\cite{rusch2009iaslc}. The delineation of \ac{LNS} in the current clinical workflow is predominantly a manual process using \ac{CT} images. Visual assessment and manual delineation is a challenging and time-consuming task even for experienced physicians, since converting text definitions of \ac{IASLC} to precise 3D voxel-wise annotations can be error prone leading to large intra- and inter-user variability~\cite{chapet2005ct}.

\begin{figure}[t]
\centering
\includegraphics[width=0.85\textwidth]{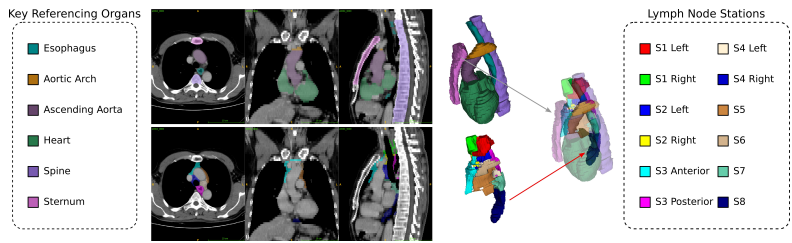}
\caption{An illustration of \ac{LNS} and key referencing organs. The top row illustrates the auto-searched top-6 key referencing organs; the bottom row depicts the 12 \acp{LNS}.} \label{Fig:LNS_demo}
\end{figure}

Deep \acp{CNN} have made remarkable progress in segmenting organs and tumors in medical imaging~\cite{tang2019clinically,zhang2020robust,jin2019accurate,jin2019deep,guo2020organ,jin2020deeptarget}. Only a handful of non-deep learning studies have tackled the automated LNS segmentation~\cite{feuerstein2012mediastinal,matsumoto2014automatic,sarrut2014learning,liu2016mediastinal}. A LNS atlas was established using deformable registration~\cite{feuerstein2012mediastinal}. Predefined margins from manually selected organs, such as the aorta, trachea, and vessels, were applied to infer \acp{LNS}~\cite{liu2016mediastinal}, which was not able to accurately adapt to individual subject. Other methods~\cite{matsumoto2014automatic,sarrut2014learning} built fuzzy models to directly parse the LNS or learn the relative positions between LNS and some referencing organs. Average location errors ranging from $6.9$mm to $34.2$mm were reported using 22 test cases in~\cite{matsumoto2014automatic}, while an average Dice score (DSC) of $66.0\%$ for $10$ LNSs in 5 patients was observed in~\cite{sarrut2014learning}.

In this work, we propose the DeepStationing -- an anatomical context encoded deep \ac{LNS} parsing framework with key organ auto-search. We first segment a comprehensive set of 22 chest organs related to the description of LNS according to \ac{IASLC} guideline. As inspired by~\cite{guo2020organ}, the 22 organs are stratified into the anchor or non-anchor categories. The predictions of the former category are exploited to guide and boost the segmentation performance of the later category. Next, \ac{CT} image and referencing organ predictions are combined as different input channels to the \ac{LNS} parsing module. The 22 referencing organs are identified by human experts. However, relevant but different from the human process, \ac{CNN} may require a particular set of referencing organs (key organs) that can opt for optimal performance. Therefore, we automatically search for the key organs by applying a channel-weighting to the input organ prediction channels based on differentiable neural search~\cite{liu2018darts}. The auto-searched final top-6 key organs, i.e., esophagus, aortic arch, ascending aorta, heart, spine and sternum (shown in \Fig~\ref{Fig:LNS_demo}), facilitate our DeepStationing method to achieve high LNS parsing accuracy. We adopt 3D nnU-Net~\cite{isensee2020nnu} as our segmentation and parsing backbone. Extensive 4-fold cross-validation is conducted using a dataset of $98$ \ac{CT} images with $12$ \ac{LNS} + $22$ Organ labels each, as \textit{the first of its kind} to date. Experimental results demonstrate that deep model encoded with the spatial context of auto-searched key organs significantly improves the LNS paring performance, resulting in an average \ac{DS} of $81.1\%\pm6.1\%$, which is $5.0\%$ and $19.2\%$ higher over the pure CT-based deep model and the most recent relevant work~\cite{liu2016mediastinal} (from our re-implementations), respectively.

\section{Method}
\begin{figure}[t]
\centering
\includegraphics[width=0.85\textwidth]{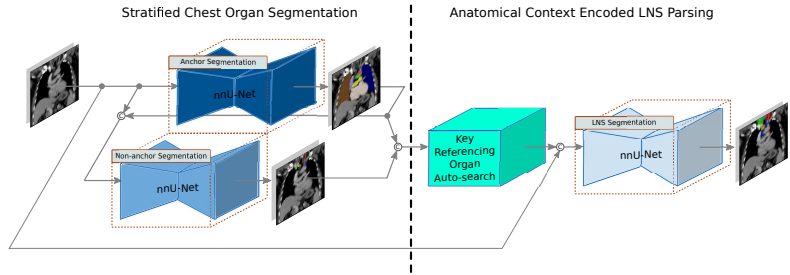}
\caption{Overall workflow of our DeepStationing, which consists of stratified chest organ segmentation and anatomical context encoded \ac{LNS} parsing with key organ auto-search.  }\label{Fig:LNS_overall_pipeline}
\end{figure}

\Fig~\ref{Fig:LNS_overall_pipeline} depicts the overview of our DeepStationing framework, consisting of two major modularized components: (1) stratified chest organ segmentation; (2) context encoded \ac{LNS} parsing with key organ auto-search. 

\subsection{Stratified Chest Organ Segmentation}\label{sec:prior_seg}
To provide the spatial context for LNS parsing, we first segment a comprehensive set of 22 chest organs related to the description of LNS. Simultaneously segmenting a large number of organs increase optimization difficulty leading to sub-optimal performance. Motivated by ~\cite{guo2020organ}, we stratify 22 chest organs into the anchor and non-anchor categories. Anchor organs have high contrast, hence, it is relatively easy and robust to segment them directly using the deep appearance features. Anchor organs are first segmented, and their results serve as ideal candidates to support the segmentation of other difficult non-anchors. We use two CNN branches to stratify the anchor and non-anchor organ segmentation. With predicted anchor organs as additional input, the non-anchor organs are segmented. Assuming $N$ data instances, we denote the training data as $\mathbb{S}=\left\{ X_n, Y_n^{\mathrm{A}}, Y_n^{\mathrm{\neg A}}, Y_n^{\mathrm{L}}, \right\} _{n=1}^{N}$, where $X_n$, $Y_n^{\mathrm{A}}$, $Y_n^{\mathrm{\neg A}}$ and $Y_n^{\mathrm{L}}$ denote the input \ac{CT} and ground-truth masks for the anchor, non-anchor organs and \ac{LNS}, respectively. Assuming there are $C_{\mathrm{A}}$ and $C_{\mathrm{\neg A}}$ classes for anchor and non-anchor organs and dropping $n$ for clarity, our organ segmentation module generate the anchor and non-anchor organ predictions at every voxel location, $j$, and every output class, $c$:
    \begin{align}
        \hat{Y}^{\mathrm{A}}_c(j) = p^{\mathrm{A}}\left( Y^{\mathrm{A}}(j) = c\, |\, X ; \mathbf{W}^{\mathrm{A}}\right)  \mathrm{,} & \quad \hat{\mathbf{Y}}^{\mathrm{A}}=\left[ \hat{Y}^{\mathrm{A}}_1\ldots\hat{Y}^{\mathrm{A}}_{C_{\mathrm{A}}} \right]  \label{eq:anchor} \mathrm{,} \\
        \hat{Y}^{\mathrm{\neg A}}_c(j) = p^{\mathrm{\neg A}}\left( Y^{\mathrm{\neg A}}(j) = c\, |\, X,  \hat{\mathbf{Y}}^{\mathrm{A}};  \mathbf{W}^{\mathrm{\neg A}}\right) \mathrm{,} & \quad
        \hat{\mathbf{Y}}^{\mathrm{\neg A}}=\left[ \hat{Y}^{\mathrm{\neg A}}_1\ldots\hat{Y}^{\mathrm{\neg A}}_{C_{\mathrm{\neg A}}} \right] \label{eq:non-anchor} \mathrm{,}
    \end{align}
where $p^{(\ast)}(.)$ denotes the \ac{CNN} functions and and $\hat{Y}^{(\ast)}_c$ for the output segmentation maps. Here, we combine both anchor and non-anchor organ predictions into an overall prediction map $\hat{\mathbf{Y}}^{\mathfrak{A}}=\hat{\mathbf{Y}}^{\mathrm{A}} \cup \hat{\mathbf{Y}}^{\mathrm{\neg A}}$. Predictions are vector valued 3D masks as they provide a pseudo-probability for every class.  $\mathbf{W}^{(\ast)}$ represents the corresponding  \ac{CNN} parameters.

\subsection{Anatomical Context Encoded LNS Parsing}\label{Sec:LNS_parse}
Segmenting \ac{LNS} by only CT appearance can be error prone, since LNS highly relies on the spatial context of adjacent anatomical structures. Emulating the clinical practice of \ac{IASLC} guidelines, we incorporate the referencing organs into the training process of \ac{LNS} parsing. Given $C_{\mathrm{L}}$ classes of the \acp{LNS}, as illustrated in \Fig~\ref{Fig:LNS_overall_pipeline}, we combine the above organ predictions with \ac{CT} images to create a multi-channel input: $\left[ X, \,\, \hat{\mathbf{Y}}^{\mathfrak{A}} \right]$:
    \begin{equation}
        \hat{Y}^{\mathrm{L}}_c(j) = p^{\mathrm{L}}\left( Y^{\mathrm{L}}(j) = c \, | \, X,  \hat{\mathbf{Y}}^{\mathfrak{A}}; \mathbf{W}^{\mathrm{L}}\right) \mathrm{,} \quad \hat{\mathbf{Y}}^{\mathrm{L}} = \left[ \hat{Y}^{\mathrm{L}}_1\ldots\hat{Y}^{\mathrm{L}}_{C_{\mathrm{L}}} \right] \mathrm{.}
    \end{equation}
Thereupon, the LNS parsing module leverages both the CT appearance and the predicted anatomical structures, implicitly encoding the spatial distributions of referencing organs during training. Similar to Eq.~\eqref{eq:anchor}, we have the \ac{LNS} prediction in its vector-valued form as $\hat{\mathbf{Y}}^{\mathrm{L}}$. 

\subsubsection{Key Organ Auto-search}\label{Sec:organ_search}
The 22 referencing organs are previously selected according to the IASLC guideline. Nevertheless for deep learning based LNS model training, those manually selected organs might not lead to the optimal performance. Considering the potential variations in organ location and size distributions, and differences in automated organ segmentation accuracy, we hypothesize that the deep \ac{LNS} parsing model would benefit from an automated reference organ selection process that are tailored to this purpose. Hence, we use the differentiable neural search~\cite{guo2020organ} to search the key organs by applying a channel-weighting strategy to input organ masks. We make the search space continuous by relaxing the selection of the referencing organs to a Softmax function over the channel weights of the one-hot organ predictions $\hat{\mathbf{Y}}^{\mathfrak{A}}$. For $C_{\mathrm{L}}$ classes, we define a set of $C_{\mathrm{L}}$ learn-able logits for each channel, denoted as $\alpha_c, \forall c \in\left[1\cdots C_{\mathrm{L}}\right]$. The channel weight $\phi_c$ for a referencing organ is defined as:
    \begin{align}
        \phi_c = \dfrac{\text{exp}\left( \alpha_{c} \right)}{\sum_{m=1}^{C_{\mathrm{L}}}\text{exp}\left( \alpha_{m} \right)} &  \mathrm{,}  \quad \Phi = \left[\phi_1 \cdots \phi_{C_{\mathrm{L}}} \right] \mathrm{,} \\
        F(\hat{Y}^{\mathfrak{A}}_c, \phi_c) =  \phi_c \cdot \hat{Y}^{\mathfrak{A}}_c &   \mathrm{,}  \quad F (\hat{\mathbf{Y}}^{\mathfrak{A}}, \Phi) = \left[F(\hat{Y}^{\mathfrak{A}}_1, \phi_1) \cdots F(\hat{Y}^{\mathfrak{A}}_{C_{\mathrm{L}}}, \phi_{C_{\mathrm{L}}}) \right]
    \end{align}
where $\Phi$ denotes the set of channel weights and $F(\phi_c, \hat{Y}^{\mathfrak{A}}_c)$ denotes the channel-wise multiplication between the scalar $\phi_c$ and the organ prediction $\hat{Y}^{\mathfrak{A}}_c$. The input of \ac{LNS} parsing model becomes $\left[ X, \,\, F (\hat{\mathbf{Y}}^{\mathfrak{A}}, \Phi) \right]$. As the results of the key organ auto-search, we select the organs with the top-$n$ weights to be the searched $n$ key organs. In this paper, we heuristically select the $n=6$ based on the experimental results. Last, we train the \ac{LNS} parsing model using the combination of original \ac{CT} images and the auto-selected top-$6$ key organs' segmentation predictions.

\section{Experimental Results}
\noindent{\bf Dataset.} We collected $98$ contrast-enhanced venous-phase \ac{CT} images of patients with esophageal cancers underwent surgery and/or radiotherapy treatments. A board-certified radiation oncologist with 15 years of experience annotated each patient with 3D masks of $12$ \acp{LNS}, involved \acp{LN} (if any), and $22$ referencing organs related to LNS according to \ac{IASLC} guideline. The 12 annotated \ac{LN} stations are: S1 \textit{(left + right)}, S2 \textit{(left + right)}, S3 \textit{(anterior + posterior)}, S4 \textit{(left + right)}, S5, S6, S7, S8. The average \ac{CT} image size is $512 \times 512 \times 80$ voxels with an average resolution of  $0.7 \times 0.7 \times 5.0$mm. Extensive four-fold cross-validation (CV), separated at the patient level, was conducted. We report the segmentation performance using \ac{DS} in percentage, \ac{HD} and \ac{ASD} in mm. 

\begin{table}[!ht]
\caption{Mean DSCs, HDs, and ASDs, and their standard deviations of LNS parsing performance using: (1) only CT appearance; (2) CT$+$all 22 referencing organ ground-truth masks; (3) CT$+$all 22 referencing organ predicted masks; (4) CT$+$auto-searched 6 referencing organ predicted masks. The best performance scores are shown in {\bf bold}.} \label{tab: quant}
\centering
\scalebox{.85}{
\setlength{\tabcolsep}{4.5mm}{
\begin{tabular}{|l|r|r|r|r|}
\hline
\multicolumn{1}{|c|}{} & \multicolumn{1}{c|}{} & \multicolumn{1}{c|}{} & \multicolumn{1}{c|}{} & \multicolumn{1}{c|}{} \\
\multicolumn{1}{|c|}{\multirow{-2}{*}{LNS}} & \multicolumn{1}{c|}{\multirow{-2}{*}{CT Only}} & \multicolumn{1}{c|}{\multirow{-2}{*}{\begin{tabular}[c]{@{}c@{}}+22 \\ Organ GT\end{tabular}}} & \multicolumn{1}{c|}{\multirow{-2}{*}{\begin{tabular}[c]{@{}c@{}}+22 \\ Organ Pred\end{tabular}}} & \multicolumn{1}{c|}{\multirow{-2}{*}{\begin{tabular}[c]{@{}c@{}}+6 Searched \\ Organ Pred\end{tabular}}} \\ \hline
\multicolumn{5}{|c|}{\cellcolor[HTML]{EFEFEF}DSC} \\ \hline
S1 Left & 78.1 $\pm$ 6.8 & 84.3 $\pm$ 4.5 & 82.3 $\pm$ 4.6 & \textbf{85.1 $\pm$ 4.0} \\
S1 Right & 76.8 $\pm$ 5.0 & 84.3 $\pm$ 3.4 & 82.2 $\pm$ 3.4 & \textbf{85.0 $\pm$ 4.1} \\
S2 Left & 66.9 $\pm$ 11.4 & 75.8 $\pm$ 9.0 & 73.7 $\pm$ 8.9 & \textbf{76.1 $\pm$ 8.2} \\
S2 Right & 70.7 $\pm$ 8.5 & 74.8 $\pm$ 7.6 & 72.8 $\pm$ 7.6 & \textbf{77.5 $\pm$ 6.4} \\
S3 Anterior & 77.4 $\pm$ 4.9 & 79.8 $\pm$ 5.6 & 79.7 $\pm$ 5.6 & \textbf{81.5 $\pm$ 4.9} \\
S3 Posterior & 84.6 $\pm$ 3.1 & 87.9 $\pm$ 2.8 & 87.8 $\pm$ 2.9 & \textbf{88.6 $\pm$ 2.7} \\
S4 Left & 74.1 $\pm$ 8.2 & 77.0 $\pm$ 8.9 & 76.9 $\pm$ 8.9 & \textbf{77.9 $\pm$ 9.4} \\
S4 Right & 73.8 $\pm$ 8.9 & 74.9 $\pm$ 9.3 & 74.9 $\pm$ 9.4 & \textbf{76.7 $\pm$ 8.3} \\
S5 & 72.6 $\pm$ 6.7 & 73.2 $\pm$ 7.4 & 73.2 $\pm$ 7.4 & \textbf{77.9 $\pm$ 8.0} \\
S6 & 72.4 $\pm$ 5.7 & 74.9 $\pm$ 4.4 & 74.8 $\pm$ 4.5 & \textbf{75.7 $\pm$ 4.3} \\
S7 & 85.0 $\pm$ 5.1 & 86.6 $\pm$ 5.8 & 86.6 $\pm$ 5.8 & \textbf{88.0 $\pm$ 6.1} \\
S8 & 80.9 $\pm$ 6.1 & 84.0 $\pm$ 5.9 & 82.0 $\pm$ 5.9 & \textbf{84.3 $\pm$ 6.3} \\ \hdashline
Average & 76.1 $\pm$ 6.7 & 79.8 $\pm$ 6.2 & 78.9 $\pm$ 6.3 & \textbf{81.1 $\pm$ 6.1} \\ \hline
\multicolumn{5}{|c|}{\cellcolor[HTML]{EFEFEF}HD} \\ \hline
S1 Left & 11.9 $\pm$ 3.2 & 12.3 $\pm$ 6.0 & 27.6 $\pm$ 38.8 & \textbf{10.3 $\pm$ 4.1} \\
S1 Right & 18.0 $\pm$ 29.3 & 10.6 $\pm$ 2.6 & 61.1 $\pm$ 97.6 & \textbf{9.7 $\pm$ 1.8} \\
S2 Left & 13.3 $\pm$ 9.2 & 9.7 $\pm$ 3.1 & 35.6 $\pm$ 76.9 & \textbf{9.2 $\pm$ 3.1} \\
S2 Right & 36.3 $\pm$ 61.7 & 10.8 $\pm$ 3.0 & 10.8 $\pm$ 3.0 & \textbf{9.5 $\pm$ 3.2} \\
S3 Anterior & 41.7 $\pm$ 62.4 & 13.5 $\pm$ 4.9 & 50.4 $\pm$ 79.1 & \textbf{12.2 $\pm$ 4.3} \\
S3 Posterior & 9.1 $\pm$ 3.3 & 8.0 $\pm$ 2.0 & 18.0 $\pm$ 30.9 & \textbf{7.6 $\pm$ 1.9} \\
S4 Left & 11.5 $\pm$ 4.9 & 14.7 $\pm$ 22.2 & 14.5 $\pm$ 22.2 & \textbf{9.8 $\pm$ 3.8} \\
S4 Right & 32.8 $\pm$ 69.7 & \textbf{9.8 $\pm$ 3.5} & 16.2 $\pm$ 21.5 & \textbf{9.8 $\pm$ 3.6} \\
S5 & 36.4 $\pm$ 56.4 & 20.5 $\pm$ 35.2 & 38.1 $\pm$ 60.3 & \textbf{10.9 $\pm$ 4.0} \\
S6 & 19.2 $\pm$ 30.6 & 8.6 $\pm$ 2.5 & 52.5 $\pm$ 85.3 & \textbf{8.5 $\pm$ 2.7} \\
S7 & 26.3 $\pm$ 42.6 & 9.6 $\pm$ 3.7 & 9.6 $\pm$ 3.7 & \textbf{9.5 $\pm$ 3.5} \\
S8 & 14.5 $\pm$ 6.0 & 13.6 $\pm$ 5.7 & 13.1 $\pm$ 5.8 & \textbf{12.2 $\pm$ 6.2} \\ \hdashline
Average & 22.6 $\pm$ 31.6 & 11.8 $\pm$ 7.9 & 28.9 $\pm$ 43.8 & \textbf{9.9 $\pm$ 3.5} \\ \hline
\multicolumn{5}{|c|}{\cellcolor[HTML]{EFEFEF}ASD} \\ \hline
S1 Left & 1.6 $\pm$ 0.8 & 1.3 $\pm$ 0.6 & 1.4 $\pm$ 1.0 & \textbf{0.9 $\pm$ 0.5} \\
S1 Right & 1.8 $\pm$ 0.8 & 1.2 $\pm$ 0.5 & 1.6 $\pm$ 1.1 & \textbf{0.9 $\pm$ 0.5} \\
S2 Left & 1.4 $\pm$ 0.8 & 1.0 $\pm$ 0.6 & 1.3 $\pm$ 0.8 & \textbf{0.8 $\pm$ 0.6} \\
S2 Right & 1.5 $\pm$ 0.8 & 1.3 $\pm$ 0.7 & 1.3 $\pm$ 0.7 & \textbf{1.0 $\pm$ 0.7} \\
S3 Anterior & 1.0 $\pm$ 0.8 & 0.7 $\pm$ 0.4 & 0.9 $\pm$ 0.9 & \textbf{0.6 $\pm$ 0.4} \\
S3 Posterior & 0.9 $\pm$ 0.5 & \textbf{0.6 $\pm$ 0.3} & 0.8 $\pm$ 1.1 & \textbf{0.6 $\pm$ 0.4} \\
S4 Left & 1.0 $\pm$ 0.6 & 1.4 $\pm$ 2.7 & 1.2 $\pm$ 1.6 & \textbf{0.8 $\pm$ 0.6} \\
S4 Right & 1.5 $\pm$ 1.0 & 1.4 $\pm$ 1.0 & 1.5 $\pm$ 1.0 & \textbf{1.3 $\pm$ 1.0} \\
S5 & 1.3 $\pm$ 0.6 & 1.9 $\pm$ 3.4 & 1.6 $\pm$ 1.8 & \textbf{1.0 $\pm$ 0.5} \\
S6 & 0.8 $\pm$ 0.4 & 0.7 $\pm$ 0.3 & 1.0 $\pm$ 1.1 & \textbf{0.6 $\pm$ 0.3} \\
S7 & 0.9 $\pm$ 0.7 & 0.8 $\pm$ 0.6 & 0.8 $\pm$ 0.6 & \textbf{0.7 $\pm$ 0.6} \\
S8 & 1.7 $\pm$ 1.2 & 1.6 $\pm$ 1.1 & 1.6 $\pm$ 1.1 & \textbf{1.3 $\pm$ 1.3} \\ \hdashline
Average & 1.3 $\pm$ 0.7 & 1.1 $\pm$ 1.0 & 1.3 $\pm$ 1.1 & \textbf{0.9 $\pm$ 0.6} \\ \hline
\end{tabular}
}} 
\end{table}

\noindent{\bf Implementation details.} We adopt the nnU-Net~\cite{isensee2020nnu} with DSC+CE losses as our backbone for all experiments due to its high accuracy on many medical image segmentation tasks. The nnU-Net has been proposed to automatically adapt different preprocessing strategies (i.e., the training image patch size, resolution, and learning rate) to a given 3D medical imaging dataset. We use the default nnU-Net settings for our model training. The total training epochs is 1000. For the organ auto-search parameter $\alpha_c$, we first fix the $\alpha_c$ for $200$ epochs and alternatively update the $\alpha_c$ and the network weights for another $800$ epochs. The rest settings are the same as the default nnU-Net setup. We implemented our DeepStationing method in PyTorch, and an NVIDIA Quadro RTX 8000 was used for training. The average training/inference time is 2.5 GPU days or 3 mins.

\subsubsection*{Quantitative Results.}\label{Sec:Eva}
We first evaluate the performance of our stratified referencing organ segmentation. The average DSC, HD and ASD for anchor and nonanchor organs are $90.0\pm4.3\%$, $16.0\pm18.0mm$, $1.2\pm1.1mm$, and $82.1\pm6.0\%$, $19.4\pm15.0mm$, $1.2\pm1.4mm$, respectively. We also train a model by segmenting all organs using only one nnUNet. The average DSCs of the anchor, non-anchor, and all organs are $86.4\pm5.1\%$, $72.7\pm8.7\%$, and $80.8\pm7.06\%$, which are $3.6\%$, $9.4\%$, and $5.7\%$ less than the stratified version, respectively. The stratified organ segmentation demonstrates high accuracy, which provides robust organ predictions for the subsequent LNS parsing model. 

\begin{figure}[!ht]
\centering
\includegraphics[width=0.8\textwidth]{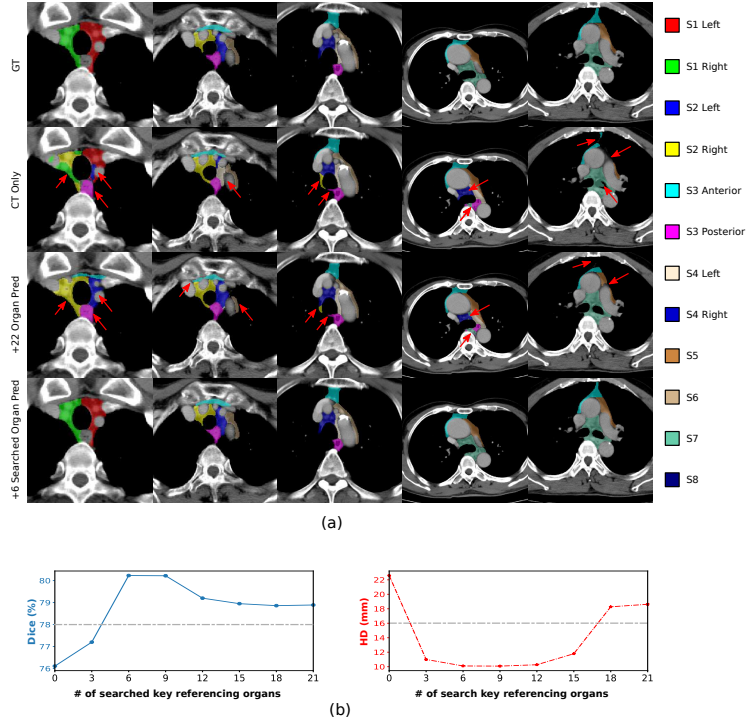}
\caption{(a) Examples of \ac{LNS} parsing results using different setups. For better comparison, red arrows are used to depict visual improvements. (b) The bottom charts demonstrate the performance using different numbers of searched referencing organs.} \label{Fig:quali}  
\end{figure}

Table~\ref{tab: quant} outlines the quantitative comparisons on different deep \ac{LNS} parsing setups. Columns 1 to 3 show the results using: 1) only \ac{CT} images, 2) \ac{CT} $+$ all $22$ ground-truth organ masks, and 3) \ac{CT} $+$ all $22$ predicted organ masks. Using only \ac{CT} images, \ac{LNS} parsing exhibits lowest performance with an average \ac{DS} of $76.1\%$ and \ac{HD} of $22.6$mm. E.g., distant false predictions is observed in the first image $2^{nd}$ row of~\Fig~\ref{Fig:quali} and false-positive S3 posterior is predicted (in pink) between the S1 and S2. When adding $22$ ground-truth organ masks as spatial context, both \ac{DS} and \ac{HD} show remarked improvements: from $76.1\%$ to $79.8\%$ in \ac{DS} and $22.6$mm to $11.8$mm in \ac{HD}. This verifies the importance and effectiveness of referencing organs in inferring LNS boundaries. However, when predicted masks of the 22 organs are used (the real testing condition), it has a significant increase in HD from $11.8$mm to $28.9$mm as compared to that using ground truth organ masks. This shows the necessity to select the key organs suited for the deep parsing model.  Finally, using the top-6 auto-searched referencing organs, our DeepStationing model achieves the best performance reaching {\bf 81.1 $\pm$ 6.1\%} DSC, {\bf 9.9 $\pm$ 3.5mm} HD and {\bf 0.9 $\pm$ 0.6mm} ASD. Qualitative examples are shown in~\Fig~\ref{Fig:quali} illustrating these performance improvements.

We auto-search for the organs that are tailored to optimize the \ac{LNS} parsing performance. Using an interval of 3, we train 7 additional \ac{LNS} parsing models, by including the top-3 up to top-21 organs. The auto-searched ranking of the 22 organs is listed as follows: \textit{esophagus, aortic arch, ascending aorta, heart, spine, sternum, V.BCV (R+L), V.pulmonary, descending aorta, V.IJV (R+L), A.CCA (R+L), V.SVC, A.pulmonary, V.azygos, bronchus (R+L), lung (R+L), trachea}, where \textit{`A'} and \textit{`V'} denote the \textit{Artery} and \textit{Vein}. The quantitative \ac{LNS} parsing results in selecting the top-n organs are illustrated in the bottom charts of \Fig~\ref{Fig:quali}. With more organs included gradually, the \ac{DS} first improves, then slightly drops after having more than top-6 organs. The performance later witnesses a sharp drop after including more than top-9 organs, then becoming steady when we include more than top-15 organs. This demonstrates that deep LNS paring model does not need a complete set of referencing organs to capture the LNS boundaries.  We choose the top-6 as our final key organs based on experimental results. We notice that the trachea, lungs, and bronchus are surprisingly ranked in the bottom-5 of the auto-search, although previous works~\cite{lu2011automatic,liu2016mediastinal} manually selected them for the LNS parsing. The assumed reasons are that those organs are usually filled with air and have clear boundaries while \ac{LNS} does not include air or air-filled organs. With the help of the other found key organs, it is relatively straightforward for the \ac{LNS} parsing \ac{CNN} to distinguish them and reject the false-positives located in those air-filled organs. We further include 6 ablation studies and segment LNS using: (1) randomly selected 6 organs; (2) top-6 organs with best organ segmentation accuracy; (3) anchor organs; (4) recommended 6 organs from the senior oncologists; (5) searched 6 organs predictions from less accurate non-stratified organ segmentor; (6) searched 6 organs GT. The randomly selected 6 organs are: \textit{V.BCV (L)}, \textit{V.pulmonary}, \textit{V.IJV (R)}, \textit{heart}, \textit{spine}, \textit{trachea}; The 6 organs with the best segmentation accuracy are: \textit{lungs (R+L)}, \textit{descending aorta}, \textit{heart}, \textit{trachea}, \textit{spine}; Oncologists recommended 6 organs are: \textit{trachea}, \textit{aortic arch}, \textit{spine}, \textit{lungs (R+L)}, \textit{descending aorta}; The DSCs for setups (1-6) are 77.2\%, 78.2\%, 78.6\%, 79.0\%, 80.2\%, 81.7\%; the HDs are 19.3mm, 11.8mm, 12.4mm, 11.0mm, 10.1mm, 8.6mm, respectively. In comparison to the LNS predictions using only CT images, the ablation studies demonstrate that using the referencing organ for LNS segmentation is the key contributor for the performance gain, and the selection and the quality of supporting organs are the main factors for the performance boost, e.g., our main results of the setups (5) and (6) show that better searched-organ delineation can help get superior LNS segmentation performance.

\noindent{\bf Comparison to previous work.} We compare the DeepStationing to the previous most relevant approach~\cite{liu2016mediastinal} that exploits heuristically pre-defined spatial margins for \ac{LNS} inference. The DeepStationing outperforms ~\cite{liu2016mediastinal} by $19.2\%$ in \ac{DS}, $30.2$mm in \ac{HD}, and $5.2$mm in \ac{ASD}. For the ease of comparison, similar to~\cite{liu2016mediastinal}, we also merge our \acp{LNS} into four \ac{LN} zones, i.e., \textit{supraclavicular} (S1), \textit{superior} (S2, S3, and S4), \textit{aortic} (S5 and S6) and \textit{inferior} (S7 and S8) zones, and calculate the accuracy of \ac{LN} instances that are correctly located in the predicted zones. DeepStationing achieves an average accuracy of $96.5\%$, or $13.3\%$ absolutely superior than \cite{liu2016mediastinal} in \ac{LN} instance counting accuracy. We tested additionally 2 backbone networks: 3D PHNN (3D UNet with a light-weighted decoding path) and 2D UNet. The DSCs of 3D PHNN and 2D UNet are 79.5\% and 78.8\%, respectively. The assumed reason for the performance drop might be the loss of the boundary precision/3D information.

\section{Conclusion}
In this paper, we propose DeepStationing as a novel framework that performs key organ auto-search based \ac{LNS} parsing on contrasted \ac{CT} images. Emulating the clinical practices, we segment the referencing organs in thoracic region and use the segmentation results to guide \ac{LNS} parsing. Different from employing the key organs directly suggested by oncologists, we search for the key organs automatically as a neural architecture search problem that can opt for optimal performance. Evaluated using a most comprehensive \ac{LNS} dataset, DeepStationing method  outperforms previous most relevant approach by a significant quantitative margin of $19.2\%$ in \ac{DS}, and is coherent to clinical explanation. This work is an important step towards reliable and automated \ac{LNS} segmentation.

\bibliographystyle{splncs04}
\bibliography{typeinst}

\end{document}